\documentclass[a4paper,12pt]{article}
\usepackage{epsfig}
\usepackage{amstex}
\usepackage{amssymb}

\newcommand{\zpc}[3]{Z.\ Phys.\ {\bf C#1} (19#2) #3}
\newcommand{\rmp}[3]{Rev.\ Mod.\ Phys.\ {\bf C#1} (19#2) #3}
\newcommand{\plb}[3]{Phys.\ Lett.\ {\bf B#1} (19#2) #3}

\newcommand{\npb}[3]{Nucl.\ Phys.\ {\bf B#1} (19#2) #3}
\newcommand{\prd}[3]{Phys.\ Rev.\ {\bf D#1} (19#2) #3}
\newcommand{\prdr}[3]{Phys.\ Rev.\ {\bf D#1} (19#2) R#3}
\newcommand{\prl}[3]{Phys.\ Rev.\ Lett.\ {\bf #1} (19#2) #3}

\def\simgt{\rlap{\lower 3.5 pt \hbox{$\mathchar \sim$}} \raise 1pt \hbox {$>$}}
\def\simlt{\rlap{\lower 3.5 pt \hbox{$\mathchar \sim$}} \raise 1pt \hbox {$<$}}

\newcommand{\beq}{\begin{equation}}
\newcommand{\eeq}{\end{equation}}
\newcommand{\bea}{\begin{eqnarray}}
\newcommand{\eea}{\end{eqnarray}}

\newcommand{\alps}{\mbox{$\alpha_{\mbox{\scriptsize s}}$}}
\newcommand{\alpsq}{\mbox{$\alpha_{\mbox{\scriptsize s}}^{2}$}}
\newcommand{\alptr}{\mbox{$\alpha_{\mbox{\scriptsize s}}^{3}$}}
\newcommand{\LQ}{L\!Q}
\newcommand{\LQb}{\overline{L\!Q}}
\newcommand{\mlq}{\mbox{$M_{\scriptstyle \LQ}$}}
\newcommand{\msq}{\mbox{$M^2_{\scriptstyle \LQ}$}}
\newcommand{\msbar}{\overline{\rm{MS}}}

\catcode`\@=11
\def\section{\@startsection{section}{1}{\z@}{3.5ex plus 1ex minus .2ex}
{2.3ex plus .2ex}{\large\bf}}
\def\thesection{\arabic{section}.}

\def\appendix{\setcounter{section}{0}
\def\thesection{Appendix \Alph{section}:}
\def\theequation{\Alph{section}.\arabic{equation}}}
\def\@citex[#1]#2{\if@filesw\immediate\write\@auxout{\string\citation{#2}}\fi
  \def\@citea{}\@cite{\@for\@citeb:=#2\do
    {\@citea\def\@citea{,\penalty\@m}\@ifundefined
       {b@\@citeb}{{\bf ?}\@warning
       {Citation `\@citeb' on page \thepage \space undefined}}%
\hbox{\csname b@\@citeb\endcsname}}}{#1}}
\def\citer{\@ifnextchar [{\@tempswatrue\@citexr}{\@tempswafalse\@citexr[]}}
\def\@citexr[#1]#2{\if@filesw\immediate\write\@auxout{\string\citation{#2}}\fi
  \def\@citea{}\@cite{\@for\@citeb:=#2\do
    {\@citea\def\@citea{--\penalty\@m}\@ifundefined
       {b@\@citeb}{{\bf ?}\@warning
       {Citation `\@citeb' on page \thepage \space undefined}}%
\hbox{\csname b@\@citeb\endcsname}}}{#1}}
\relax
\begin{document}
\thispagestyle{empty}
\begin{flushright}
RAL-97-017\\
DESY 97-063\\
CERN-TH/97-67\\
April 1997
\end{flushright}
\begin{center}
\vskip 2.5cm
\boldmath
{\large \bf Pair production of scalar leptoquarks at the Tevatron}
\unboldmath
\vglue 1.5cm
\begin{sc}
{\sc M.~Kr\"{a}mer$^a$, T.~Plehn$^b$, M.~Spira$^c$, and P.M.~Zerwas$^b$}
\vglue 0.3cm
\end{sc}
$^a${\em Rutherford Appleton Laboratory, Chilton, Didcot, OX11 0QX, UK}
\vglue 0.05cm
$^b${\em Deutsches Elektronen-Synchrotron DESY, D-22603 Hamburg, FRG}
\vglue 0.05cm
$^c${\em TH Division, CERN, CH-1211 Geneva 23, Switzerland}
\end{center}
\vglue 1.7cm

\begin{abstract} 
\noindent
We present the cross section for pair production of scalar leptoquarks
$\LQ$ in $p\bar{p}$ collisions, $p+\bar{p} \to \LQ + \LQb + X$, at the
Fermilab Tevatron in next-to-leading order QCD. Including the
higher-order corrections stabilizes the theoretical prediction and
increases the size of the cross section for
renormalization/factorization scales close to the mass of the
leptoquarks. This leads to an increase of the lower bound on the mass
of scalar leptoquarks by up to 15~GeV with respect to earlier
analyses.
\end{abstract}

\vfill
\newpage


{\bf 1.} Theoretical speculations have focussed on two different
elements to explain the recently observed surplus of \mbox{DIS} $e^+p$
events at \mbox{HERA} \cite{H1-97}: contact interactions at an
effective scale $\Lambda_{eq} \;\simgt\; 1.5$~TeV, and narrow
resonance formation at a mass scale $M\sim 200$~GeV. The resonance
interpretation is based in particular on the \mbox{H1} data which
appear to cluster in a narrow range at invariant $(eq)$ masses of
about 200~GeV. Such resonances \citer{AEGLM-97,KS-97} can be
identified with scalar squarks in supersymmetric theories with
$R$-Parity breaking or with leptoquarks in general.

Leptoquarks encompass a large variety of particles which can be
classified in multiplets according to spin, isospin and hypercharge
\cite{BRW-87}. The production of leptoquarks in positron collisions
with sea quarks requires large Yukawa couplings, which can in general
not be reconciled with bounds from rare decay processes, nor with the
early $e^-p$ scattering data from \mbox{HERA}. However, the Yukawa
couplings in positron collisions with valence quarks, $\lambda \sim
1/30$, are so small that they are not in conflict with any such
bounds. Moreover, they cannot be ruled out by present limits on $eeqq$
contact interactions from \mbox{LEP} \cite{OPAL-95} or from the
Tevatron \cite{CDF-96}. The most powerful competitor in this scenario
is the pair production of leptoquarks \cite{HP-88} at the
Tevatron\footnote{Another source of leptoquarks at the Tevatron is
  associated production $p\bar{p} \to \LQ + \bar{l} + X$ \cite{HP-88}.
  The cross section for this process depends on the leptoquark Yukawa
  coupling; it is significantly smaller than for leptoquark pair
  production \cite{BLUM-97}.}
\begin{equation}
p + \bar{p} \to \LQ + \LQb + X 
\end{equation} 
Mass bounds on vector leptoquarks, which decay solely into charged
leptons and quarks, appear significantly above 200~GeV in the analyses
of Refs.\cite{D0-97,CDF-93}. Even if the unknown anomalous couplings
of vector leptoquarks are chosen such as to minimize the cross
section, the minimum mass is 210~GeV, see \cite{BLUM-97,MONT-97}. The
present limits for scalar leptoquarks are near 200~GeV
\cite{D0-97}. The experimental bounds will be refined further by the
two Tevatron experiments in the near future. The limits for scalar
leptoquarks are particularly important since they are essentially
parameter-free. The cross sections for pair production of scalar
leptoquarks involve only the strong coupling constant, and they do not
depend on unknown Yukawa couplings.

Anticipating the refinement of the Tevatron limits on the masses of
scalar leptoquarks, a solid theoretical prediction of the production
cross section\footnote{Soft gluon corrections to the production of
  leptoquark pairs have been discussed in Ref.\cite{MM-90} [based,
  though, on erroneous Born calculations].} is mandatory, the more the
gap narrows between the Tevatron mass bounds and the \mbox{HERA} mass
estimates. Based on previous experience from the production of squark
pairs in hadron collisions \cite{BHSZ-95}, it is expected that
higher-order QCD corrections increase the production cross section
compared to the predictions at the Born level.  Experimental mass
bounds are therefore shifted upwards. Moreover, by reducing the
dependence of the cross section on spurious parameters, {\it i.e.} the
renormalization and factorization scales, the cross sections in
next-to-leading order QCD are under much better theoretical control
than the leading-order estimates. The next-to-leading order analysis
will be presented in this letter.

{\bf 2.} The basic processes for the production of leptoquark pairs at
the Tevatron are quark-antiquark annihilation and gluon-gluon fusion:
\begin{eqnarray}
\label{parton-processes}
q + \bar{q} \to \LQ + \LQb \nonumber \\
g + g       \to \LQ + \LQb 
\end{eqnarray}
Any non-pointlike structures of leptoquarks which may occur in
compositeness scenarios, are expected at scales only above $\sim
1$~TeV; in other scenarios leptoquarks are generically pointlike
particles.  The gluon-leptoquark interactions are therefore determined
by the non-abelian SU(3)$_C$ gauge symmetry of scalar QCD so that the
theoretical predictions for the pair production of scalar leptoquarks
are parameter-free.

The diagrams corresponding to the processes (\ref{parton-processes}),
are shown in Fig.\ref{Feynman-diagrams}a; the only new element of
scalar QCD is the quartic coupling between gluons and leptoquarks
which follows from the SU(3)$_C$ gauge invariance of the interaction.
The cross sections of the parton processes (\ref{parton-processes})
may be written as \cite{GM-82}
\begin{alignat}{3}
&\hat{\sigma}_{\mbox{\scriptsize LO}}[\,q\bar{q}\to\LQ+\LQb\,] 
&=&
\frac{\alpsq\pi}{\hat{s}}\,\frac{2}{27}\,\beta^3 \hfill \nonumber\\[2mm]
&\hat{\sigma}_{\mbox{\scriptsize LO}}[\,gg\to\LQ+\LQb\,] 
&=&
\frac{\alpsq\pi}{96\hat{s}}\,
\left[ \beta \left( 41 - 31 \beta^2 \right) 
       + \left( 18 \beta^2 - \beta^4 - 17 \right) 
         \log\frac{1+\beta}{1-\beta}
\right]
\end{alignat}
where $\sqrt{\hat{s}}$ is the invariant energy of the subprocess and
$\beta = \sqrt{1-4\msq/\hat{s}}$.  The cross sections coincide with
the cross sections for squark-pair production in the limit of large
gluino masses \cite{BHSZ-95,KL-82}. The quark-antiquark annihilation
is the driving mechanism at the Tevatron for large leptoquark masses.

The QCD radiative corrections to the order $\alps$ include virtual
corrections, the brems\-strah\-lung of gluons and contributions from
gluon--quark collisions. The virtual corrections can be classified in
self-energy diagrams and vertex corrections for quarks, gluons and
leptoquarks according to the rules of standard and scalar QCD, and
initial/final state interactions. In addition to gluon radiation off
all colored lines, gluons can also be emitted from scalar vertices.
Finally, the inelastic Compton process, diagram
(\ref{Feynman-diagrams}b), must be added in order $\alptr$ of the
cross section. The amplitudes have been evaluated in the Feynman
gauge. After the singularities are isolated by means of dimensional
regularization, the renormalization has been carried out in the
$\msbar$ scheme. The masses of the light quarks (u,d,s,c,b) have been
neglected while the mass parameter of the leptoquark has been defined
on-shell.  We have chosen a renormalization and factorization scheme
in which the massive particles (top quark, leptoquark) are decoupled
smoothly for momenta smaller than their mass \cite{CWZ-78}. This
implies that the heavy particles do not contribute to the evolution of
the QCD coupling and the parton densities. The computation of the
cross section for gluon emission has been performed by adopting the
phase space slicing method (see e.g.\ \cite{BKNS-89}): a cut-off
$\Delta$ has been introduced for the invariant mass of the
leptoquark-gluon system in the final state, which separates soft from
hard gluon radiation. If both contributions are added up, any $\Delta$
dependence disappears from the total cross section for $\Delta \to 0$.
The infrared singularities cancel when the emission of soft gluons is
added to the virtual corrections.  The remaining collinear
initial-state singularities are absorbed into the renormalization of
the parton densities \cite{AEM-79}, defined in the $\msbar$
factorization scheme.

The perturbative expansion of the total parton cross section can be
expressed in terms of scaling functions,
\begin{equation}
\label{sigma_ij}
\hat\sigma_{ij}(s,\msq) =
\frac{\alpsq(\mu^2)}{\msq}\,
\left[  f_{ij}^{(0)}(\eta) 
      + 4\pi\alps(\mu^2)\left\{f_{ij}^{(1)}(\eta,r_t)
+\overline{f}^{(1)}_{ij}(\eta)\ln\left(\frac{\mu^2}{\msq}\right)
\right\}\right]
\end{equation}
with $i,j=g,q,\overline{q}$ denoting the initial-state partons. For
simplicity, we have identified the renormalization scale with the
factorization scale $\mu_R = \mu_F = \mu$. The scaling functions
depend on the invariant parton energy $\sqrt{\hat{s}}$ through $\eta =
\hat{s}/4\msq - 1$ and, very mildly, on the ratio of the particle
masses $r_t = m_{\mbox{\scriptsize top}}/\mlq$.  The scaling functions
$f_{ij}^{(0,1)}$ and $\overline{f}^{(1)}_{ij}$ are displayed in
Fig.\ref{fij} for the quark-antiquark, gluon-gluon and gluon-quark
channels, respectively. The scaling functions $f_{ij}^{(1)}$ are
decomposed into a "virtual + soft" (V+S) part, and a "hard" (H)
gluon-radiation part; the $\ln^j\!\Delta~(j=1,2)$ singularities of the
(V+S) cross section are mapped into (H), cancelling the logarithms so
that these functions are independent of $\Delta$ in the limit
$\Delta\to 0$.

From Fig.\ref{fij}(a) we can infer that the next-to-leading order
corrections to the gluon-gluon channel are very important near the
threshold $\sqrt{\hat{s}} \gtrsim 2 \mlq$. The large size of the
corrections close to the threshold is partly due to gluon radiation
off the initial-state partons, which generates contributions of the
type $\beta\ln^j\!\beta~(j=1,2)$; the $\ln^2\!\beta$ terms are
universal and can be exponentiated \cite{MN-85}.  At the threshold,
$\sqrt{\hat{s}} \to 2 \mlq$, the next-to-leading order cross section
for the $gg$ initial states is non-zero: The Sommerfeld rescattering
contribution, due to the exchange of Coulomb gluons between the
leptoquark pair in the final state, gives rise to a $1/\beta$
singularity which compensates the phase space factor $\beta$. At large
parton energies, the hard coefficients $f_{gg}^{(1,H)}$ and
$f_{gq}^{(1)}$ approach plateaus, that are built up by the
flavor-excitation and gluon-splitting mechanisms. The exchange of
gluons in the $t$- and $u$- channels leads to an asymptotically
constant cross section, which is to be contrasted with the scaling
behavior $\sim 1/\hat{s}$ of the leading-order process.

The $p\bar{p}$ cross section is found by folding the parton cross
sections with the gluon and light--quark luminosities in $p\bar{p}$
collisions:
\begin{equation}
\sigma[p\bar{p}\to\LQ+\LQb+X] = \sum_{i,j=g,q(\overline{q})}\;
\int\!d\tau\frac{d{\cal{L}}^{ij}}{d\tau}\,
\hat{\sigma}_{ij}(\hat{s}=\tau{s})
\end{equation}
Due to the dominating $q\bar{q}$ luminosity for large parton momenta,
the total cross section is built up primarily by quark-antiquark
initial states for leptoquark masses $\mlq\;\simgt\; 100$~GeV.  For the
numerical analysis we adopt the CTEQ4M parametrization of the parton
densities \cite{CTEQ4}. The QCD coupling is evaluated in the $\msbar$
scheme for $n_{\mbox{\scriptsize lf}} = 5$ active flavors and
$\Lambda^{(5)} = 202$~MeV, and the top quark mass is set to
$m_{\mbox{\scriptsize top}} = 175$~GeV \cite{CDF-95}.

The scale dependence of the theoretical prediction is reduced
significantly when higher order QCD corrections are included. This is
demonstrated in Fig.\ref{scale-dependence} where we compare the
renormalization/factorization scale dependence at the leading and
next-to-leading order of the total cross section. For a consistent
comparison of the LO and NLO results, we have calculated all
quantities [{\it i.e.}\ $\alps(\mu^2)$, the parton densities, and the
partonic cross sections] in leading and next-to-leading order,
respectively.  The scale dependence of the leading-order cross section
is steep and monotonic: Changing the scale from $\mu=2\mlq$ to
$\mu=\mlq/2$, the LO cross section increases by 100\%. At
next-to-leading order the scale dependence is strongly reduced, to
about 30\% in this interval. The NLO cross section runs through a
broad maximum near $\mu\sim \mlq/2$, which supports the stable
behavior in~$\mu$.

{\bf 3.} The QCD radiative corrections enhance the cross section for
the production of leptoquarks above the central value $\mu \sim \mlq$.
If, as often done, a LO cross section is calculated in a hybrid form,
{\it i.e.} using parton cross sections in Born approximation but NLO
parton densities and $\alps$ in two-loop order, the enhancement is
close to 40\%. If, moreover, the so-defined LO cross section is
evaluated at large scales $\mu \sim \sqrt{\hat{s}}$, the enhancement
in next-to-leading order is as big as $\sim 70\%$, nearly independent
of the leptoquark mass.  The convergence of the perturbative approach
should however be judged by examining a properly defined $K$-factor,
$K = \sigma_{\mbox{\scriptsize NLO}} / \sigma_{\mbox{\scriptsize
    LO}}$, with all quantities in the numerator and denominator
calculated consistently in NLO and LO, respectively.  In the
interesting mass range between $150 \leq \mlq \leq 250$~GeV, these
K-factors vary only between 1.20 and 1.08, as shown in Table
\ref{table}. They are small enough to assure a reliable perturbative
expansion.\footnote{The results nearly coincide with the cross
  sections for the production of squark--antisquark pairs in the limit
  of large gluino mass.  Supersymmetry predicts quartic
  self-interactions to order $\alpsq$ between the squarks. No such
  self-couplings have been considered hitherto for leptoquarks in
  general. These couplings affect the cross sections only through
  rescattering corrections involving heavy leptoquark loops so that
  their impact on the cross section is small.  We have checked that
  the results after removing these diagrams are identical for large 
  gluino masses.}

The impact of the next-to-leading order QCD corrections on the present
experimental lower mass limits for scalar leptoquarks is illustrated
in Fig.\ref{mass-bounds}. We compare the NLO result based on the
default settings (CTEQ4M parton densities and $\mu=\mlq$) with the LO
cross section adopted in earlier analyses (CTEQ3M parton densities
\cite{CTEQ3} and two-loop $\alps$, evaluated at $\mu$ equal to the
partonic centre-of-mass energy $\sqrt{\hat{s}}\,$).\footnote{It is not
  legitimate to use $\mu = \sqrt{\hat{s}}$ beyond LO since this choice
  of scale results in an error of order $\alps$ no matter how
  accurately the hard scattering cross section is calculated
  \cite{CSS-89}.  By contrast, the choice $\mu=\mlq$ is legitimate
  and, moreover, provides a reasonable starting point for the
  perturbative expansion, leading to well-controlled higher-order
  corrections as evident from Fig.\ref{scale-dependence}.} Taken at
face value, the next-to-leading order corrections increase the mass
limit for a first-generation scalar leptoquark by about 15~GeV. [The
shift is much smaller if the LO cross sections are evaluated for LO
parton densities and $\alps$ at the renormalization/factorization
scale $\mu=\mlq$, as demonstrated by the broken line in
Fig.\ref{mass-bounds}.]  The shaded band reflects the remaining
theoretical uncertainty at NLO due to the choice of the
renormalization/factorization scale when $\mu$ is varied in the range
$\mlq/2 \le \mu \le 2 \mlq$.  Since the cross section in the
interesting mass region $\mlq\;\simgt\; 150$~GeV is built up mainly by
the quark-antiquark channels, thus based on well-measured parton
densities, the variation between different parton parametrizations
(CTEQ4M \cite{CTEQ4}, GRV \cite{GRV} and MRS(R2) \cite{MRS}) is less
than 5~\%.

Evaluated on the basis of the NLO cross sections presented in this
letter, the data from the Fermilab Tevatron, which were presented in
Refs.\cite{D0-97}, lead to parameter-free lower limits of about $\mlq >
190$~GeV and 210~GeV for scalar leptoquarks decaying to charged
leptons in the present \mbox{D0} and \mbox{CDF} analyses.

{\small \noindent {\bf Acknowledgements:}\\
  We have benefitted from discussions and communications with
  S.~Hagopian, S.~Lammel, and G.~Landsberg. Special thanks go to
  J.~Bl\"umlein, H.~Dreiner, M.~Mangano, and H.~Weerts for the careful
  reading and valuable comments on the manuscript.


\newpage

\renewcommand{\arraystretch}{1.15}
\begin{table}[ht]
\begin{center}
\begin{tabular}{|c||lcccc|c|} \hline
      $\mlq$ [GeV]
    &
    & $\sigma_{q\bar{q}}$
    & $\sigma_{gg}$
    & $\tilde{\sigma}_{gq}$
    & $\sigma_{\mbox{\scriptsize tot}}$ [pb]
    & K                               \\ \hline \hline
      150
    &  LO 
    &  0.741
    &  0.244
    & 
    &  0.985
    &                                 \\
    &  NLO
    &  0.722
    &  0.490
    & -0.028
    &  1.184
    &  1.20                            \\ \hline
      175
    &  LO
    &  0.318
    &  0.071
    &
    &  0.389
    &                                 \\
    &  NLO
    &  0.311
    &  0.146
    & -0.010
    &  0.447
    &  1.15                            \\ \hline
      200
    &  LO
    &  0.142
    &  0.022
    & 
    &  0.164
    &                                 \\
    &  NLO
    &  0.141
    &  0.047
    & -0.004
    &  0.184
    &  1.12                            \\ \hline
      250
    &  LO
    &  0.030
    &  0.003
    & 
    &  0.033
    &                                 \\
    &  NLO
    &  0.030
    &  0.006
    & -0.001
    &  0.035
    &  1.08                            \\ \hline
\end{tabular}
\end{center}
\caption[dummy]{\it LO and NLO results for the total cross
  section $p+\bar{p}\to\LQ+\LQb+X$ at the Tevatron energy
  $\sqrt{s} = 1.8$~TeV for various values of the leptoquark mass
  $\mlq$. All quantities [$\alps(Q^2)$, the parton densities, and the
  partonic cross sections] have been calculated consistently in leading
  and next-to-leading order, respectively. Also shown is the $K$~factor
  defined as 
  $K=\sigma_{\mbox{\scriptsize NLO}}/\sigma_{\mbox{\scriptsize LO}}$. 
  The CTEQ4M(L) parton densities with the associated values of
  $\alps$ and the central scale $\mu=\mlq$ have been adopted. 
  [The negative sign of $\tilde{\sigma}_{gq}$ is a mere artifact of 
  subtracting collinear initial-state singularities via mass 
  factorization.]}
\label{table} 
\end{table}
\renewcommand{\arraystretch}{1.0}

\newpage

\begin{figure}[ht]
\begin{center}\noindent
\epsfig{file=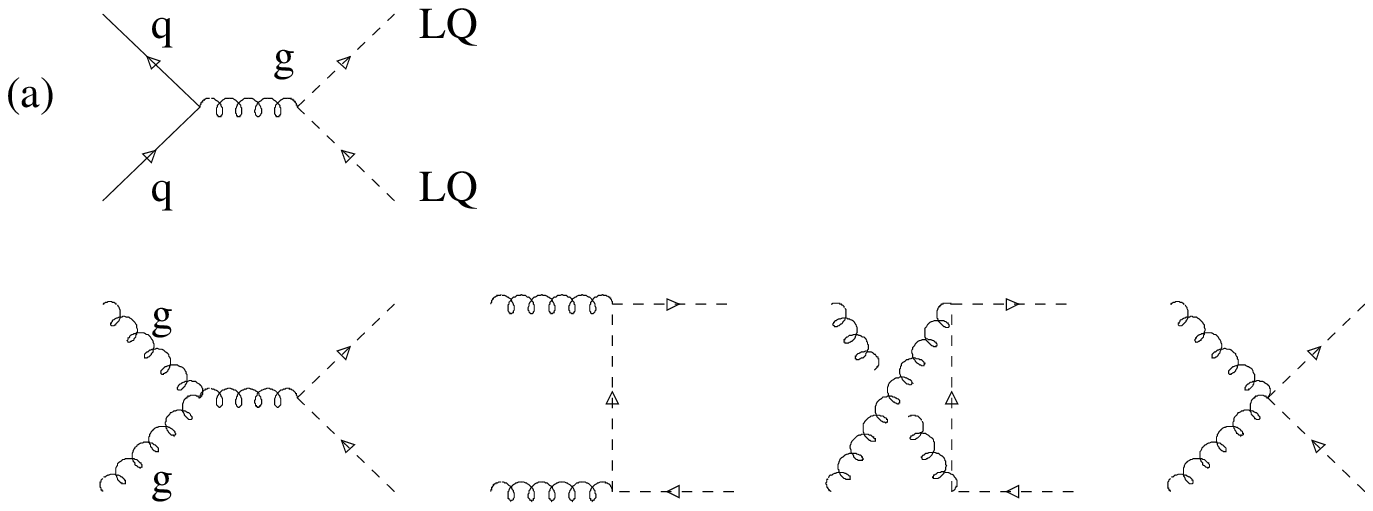,width=11.5cm} \\[2mm]
\epsfig{file=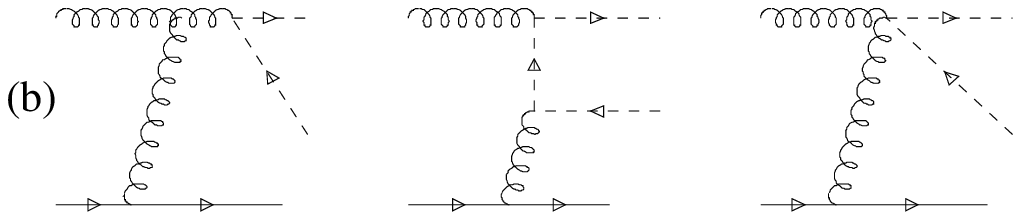,width=11.5cm}
\vspace*{-2cm}
\end{center}
\caption[dummy]{\it Generic diagrams for pair production of scalar
  leptoquarks in hadron collisions: 
  (a) $q\bar{q}$ annihilation and $gg$ fusion; 
  (b) the gluon-quark subprocess.}
\label{Feynman-diagrams}
\end{figure}

\newpage

\begin{figure}[ht]
\begin{center}
\epsfig{file=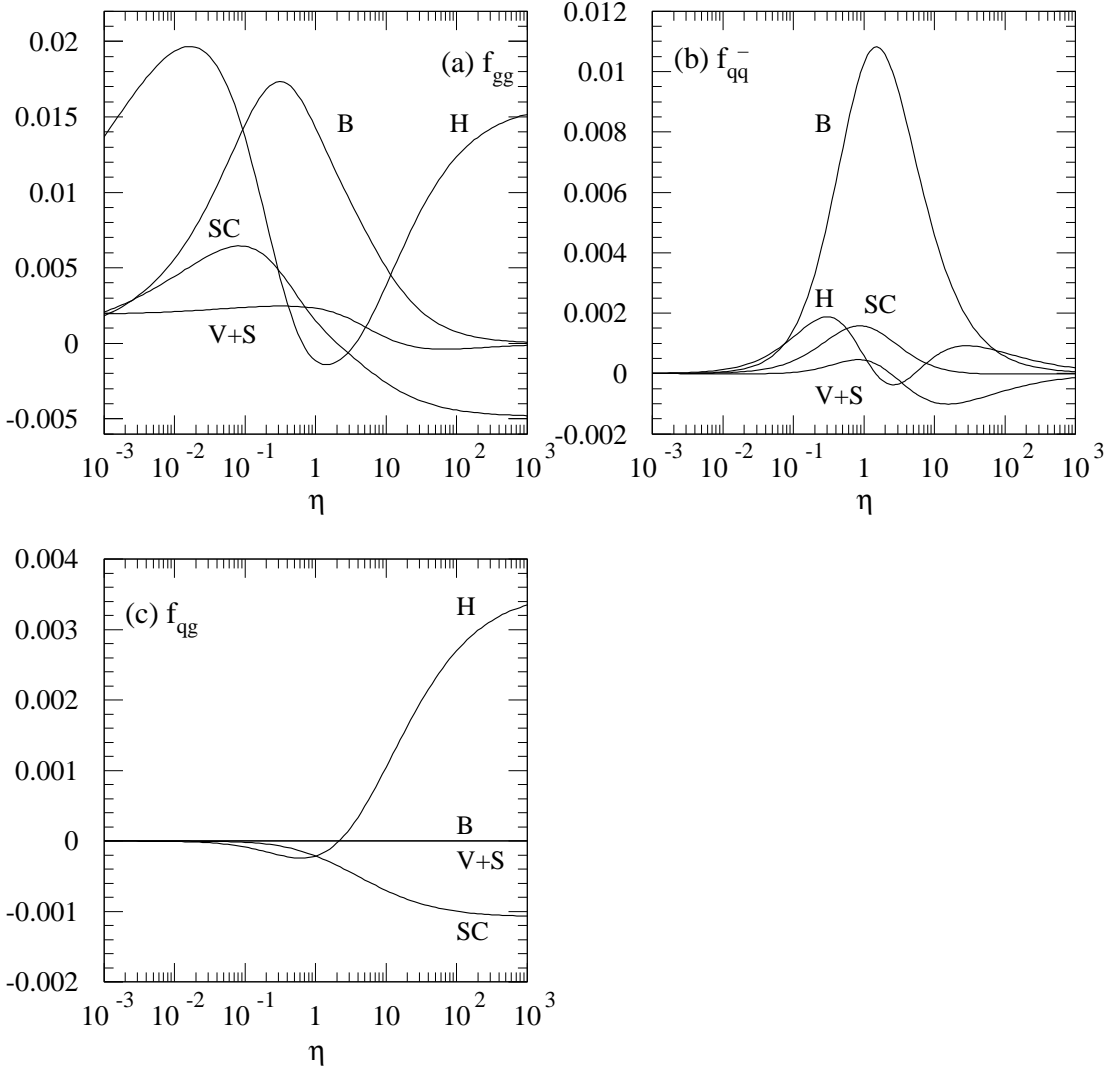,width=15cm}
\end{center}
\caption[dummy]{\it The scaling functions for scalar leptoquark
  pair production in $gg$ (a), $q\bar{q}$ (b), and $gq$ (c) collisions
  versus $\eta = \hat{s}/4\msq - 1$. The notation follows
  Eq.(\ref{sigma_ij}); B denotes the lowest-order scaling function,
  \mbox{V+S} the sum of virtual and soft corrections, H the
  contribution of hard gluon emission, \mbox{SC} the scale-dependent
  scaling function.}
\label{fij}
\end{figure}

\newpage

\begin{figure}[ht]
\begin{center}
\epsfig{file=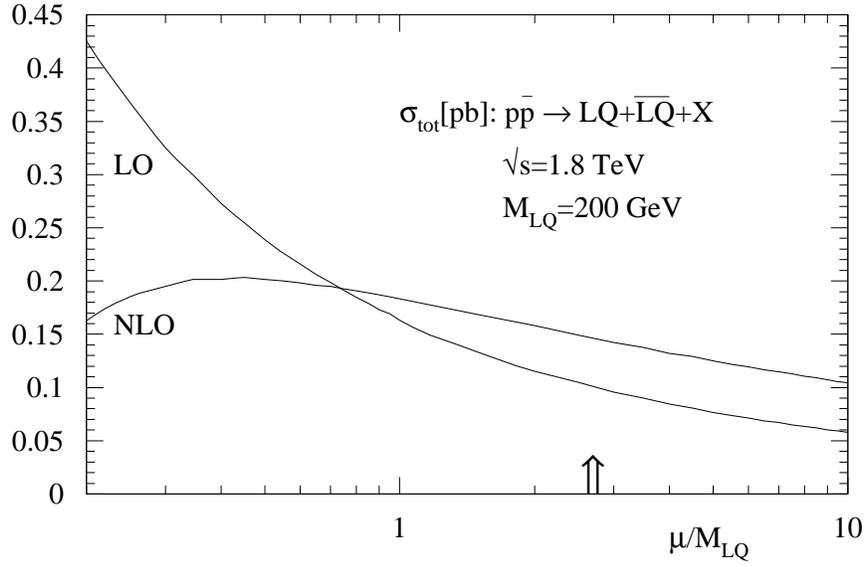,width=12cm}
\end{center}
\caption[dummy]{\it Renormalization/factorization
  scale dependence of the total cross section $p+\bar{p}\to\LQ+\LQb+X$
  at the Tevatron energy $\sqrt{s} = 1.8$~TeV.  Parameters as
  described in the text. The arrow indicates the value of the average
  invariant energy $<\!\hat{s}\!>^{1/2}$ in the hard subprocess.}
\label{scale-dependence}
\end{figure}

\newpage

\begin{figure}[ht]
\begin{center}
\epsfig{file=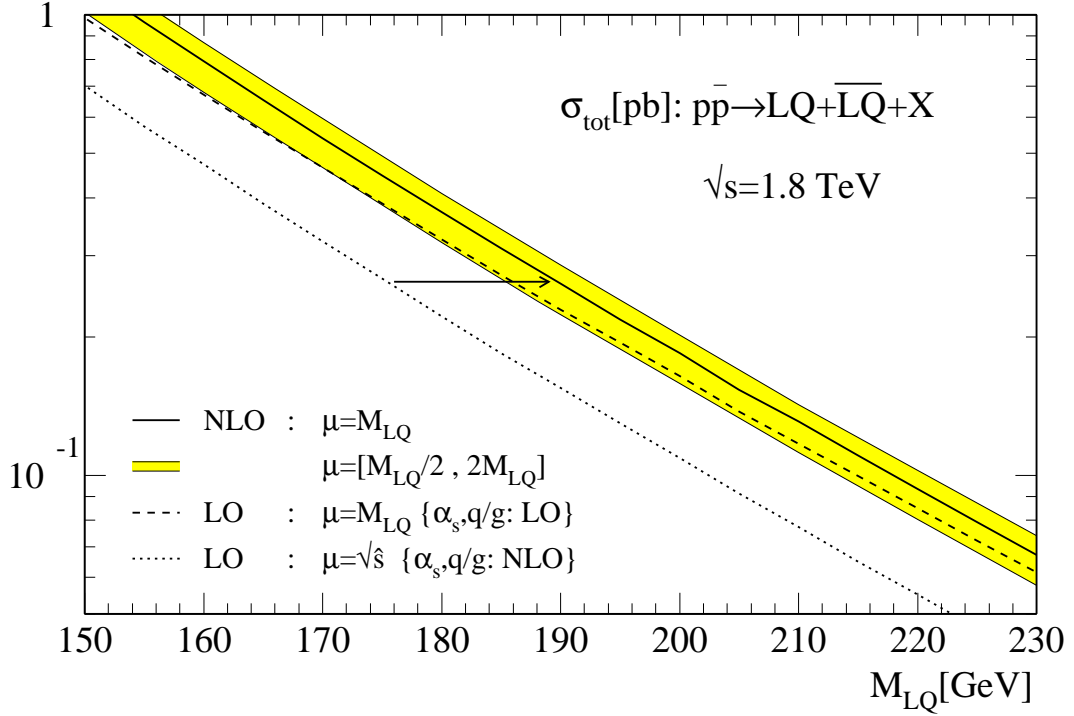,width=15cm}
\end{center}
\caption[dummy]{\it The total
  cross section for the production of leptoquark pairs,
  $p+\bar{p}\to\LQ+\LQb+X$, at the Tevatron energy $\sqrt{s} =
  1.8$~TeV as a function of the leptoquark mass $\mlq$.  The
  next-to-leading order result (NLO) is compared with various options
  of parameters in, partially hybrid, LO calculations.  The variation
  of the NLO cross section with the value of the
  renormalization/factorization scale is indicated by the shaded band.
  The increase of the lower bound on the leptoquark mass, compared to
  the earlier analyses, is demonstrated by the horizontal arrow.}
\label{mass-bounds}
\end{figure}


\end{document}